# Develop and Optimize 5DCT Imaging Simulation and Reconstruction Methods

Yuhao Wang[1], Zhendong Zhang[1], Edward Robert Criscuolo[1], John Ginn[1], Ke Lu[1], Yao Hao[2], Deshan Yang[1*]

[1]Department of Radiation Oncology, School of Medicine, Duke University
[2]Department of Radiation Oncology, School of Medicine, Washington University in Saint Louis

[*]Corresponding author: Deshan Yang, deshan.yang@duke.edu

## Abstract

**Background:** Stereotactic arrhythmia radiotherapy (STAR) requires accurate characterization of cardiac target motion, but conventional 4DCT resolves respiratory or cardiac motion separately and cannot capture their combined effect on thoracic anatomy.

**Purpose:** To develop and optimize a 5DCT (3D + cardiac phase + respiratory phase) imaging simulation and reconstruction pipeline, and to compare two sinogram-space interpolation methods for reconstructing images at arbitrary combinations of cardiac and respiratory phase.

**Methods:** Helical CT projections were simulated from the 4D XCAT phantom across a range of cardiac and respiratory motion states, with Poisson and electronic noise added. Ground-truth-matched volumes were generated at 5 cardiac phases and 10 respiratory amplitudes (50 total phase combinations). Because acquired projections are sparsely and unevenly distributed across this joint phase space, each target slice was reconstructed by interpolating rebinned sinogram rows to the target cardiac phase and respiratory amplitude, using either 2D scattered barycentric interpolation or 2D scattered local linear interpolation with a circular kernel for cardiac phase. Reconstructed volumes were compared to phantom ground truth using mean absolute error (MAE), and to conventional respiratory-gated 4DCT (r4DCT) reconstructed from the same simulated data.

**Results:** Both interpolation methods eliminated the severe axial misalignment artifacts present when helical projections were reconstructed without phase-space interpolation. Local linear interpolation achieved lower MAE than barycentric interpolation across most tested conditions, with the largest improvement at low pitch. The 5DCT pipeline also produced respiratory-only volumes with fewer residual cardiac-motion artifacts than conventional r4DCT reconstructed from the same projection data, including at standard clinical pitch (0.1).

**Conclusions:** 5DCT reconstruction using sinogram-space interpolation is feasible and can jointly resolve cardiac and respiratory motion with better accuracy than conventional 4DCT reconstruction. Local linear interpolation outperformed barycentric interpolation, and the pipeline offers a practical near-term route to improving 4DCT-based motion characterization for STAR and other thoracic radiotherapy applications.

# 1. Introduction

Stereotactic arrhythmia radiotherapy (STAR)[1] has emerged as a noninvasive treatment for refractory ventricular tachycardia[2-4]. STAR requires precise cardiorespiratory motion management to minimize healthy tissue toxicity and ensure accurate target dose delivery[4]. Cardiac targets experience two coupled motions: rapid cardiac oscillation (~1 Hz)[5], and slower respiratory-induced displacement (~0.1–0.3 Hz)[5], which are interdependent[6]. Cardiac motion during breath-hold, standard for cardiac-gated CT, differs from that during free breathing[7]. Respiratory motion is characterized using 4DCT, which bins projections into respiratory phases[8,9]. Cardiac motion is assessed via ECG-gated CT under breath-hold or cardiac 4DCT with slow table feed[10]. Neither approach captures the joint cardiorespiratory dynamics.

The absence of an imaging modality resolving both cardiac and respiratory dimensions, a 5DCT (3D space + cardiac phase + respiratory phase), is a critical gap. For STAR, millimeter-scale targeting errors determine success or failure[11-14]. Several investigators have recognized the need to study cardiorespiratory motion jointly[15,16]. The primary challenge is data scarcity: 5DCT distributes projections across a 2D phase space (cardiac × respiratory), reducing per-bin projections and causing axial slice misalignments[17]. Breathing irregularities during extended scans further degrade quality[9].

Beyond STAR, combined cardiorespiratory imaging could improve cardiac motion modeling and 4D dose accumulation for lung and breast radiotherapy. Recent 5D imaging efforts in cardiovascular MRI[18] and cone-beam CT[19] have not been extended to cardiorespiratory CT for radiotherapy.

This study developed a 5DCT imaging pipeline using the 4D XCAT phantom[20] for ground-truth volumes, simulating helical scans with clinical parameters, and reconstructing 5D volumes via single-slice rebinning[17], sinogram-space 2D interpolation using cardiac and respiratory phases as independent variables, and filtered back projection. Through evaluation across 10 respiratory and 5 cardiac phases, we validated 5DCT feasibility for joint cardiorespiratory motion analysis.

# 2. Methods

## 2.1. Dataset

We generated a 4D XCAT digital phantom[20] dataset for the 5DCT imaging study. The phantom produces ground-truth 3D volumes for arbitrary combinations of cardiac and respiratory phases. Figure 1 (A) shows a coronal frame of the generated 4D XCAT phantom. The contrast between the blood pool, coronary arteries, and the myocardium is enhanced to better evaluate the reconstruction quality. Figure 1 (B) shows part of the normalized respiratory (blue) and cardiac (red) waves, which were provided by the XCAT phantom settings, and could be used to record the respiratory and cardiac phases. The respiratory wave incorporated breathing irregularities to evaluate the robustness of our 5DCT reconstruction algorithm under clinically relevant conditions[9]. For our 5DCT reconstruction, we targeted 10 respiratory ampltudes (00% to 90%) and 5 cardiac phases (00%, 20%, 40%, 60%, and 80%), yielding 50 unique 3D volumes. The simulation followed real CT scanner geometry and acquisition protocols to ensure clinical relevance. The key scan parameters are summarized in Table 1.

| **Parameters** | **Value** |
|---|---|
| Number of frames of the 4D XCAT phantom | 2238 |
| Time per frame | 0.0219s |
| Frame size | 512x512x130 |
| Voxel size | $1\times1\times1$ mm$^3$ |
| CT scan time | 49.0s/14.7s |
| Longitudinal coverage | 119.7 mm |
| Mode | Helical |
| Cardiac period | 0.7500s |
| Respiratory curve duration* | 60s |
| Time per X-ray source rotation | 0.3s |
| Projections per X-ray source rotation | 984 |
| Total number of projections | 160689/48200 |
| Source-to-detector distance | 1085.6 mm |
| Source-to-isocenter distance | 595 mm |
| Pitch | 0.02/0.10 |
| Detector column number | 900 |
| Detector row number | 64 |
| Detector element size | $0.8\times0.8$ mm$^2$ |

*Table 1. Parameters of the XCAT phantom dataset and the virtual CT scan. *The respiratory motion is irregular so there is no fixed respiratory period.*

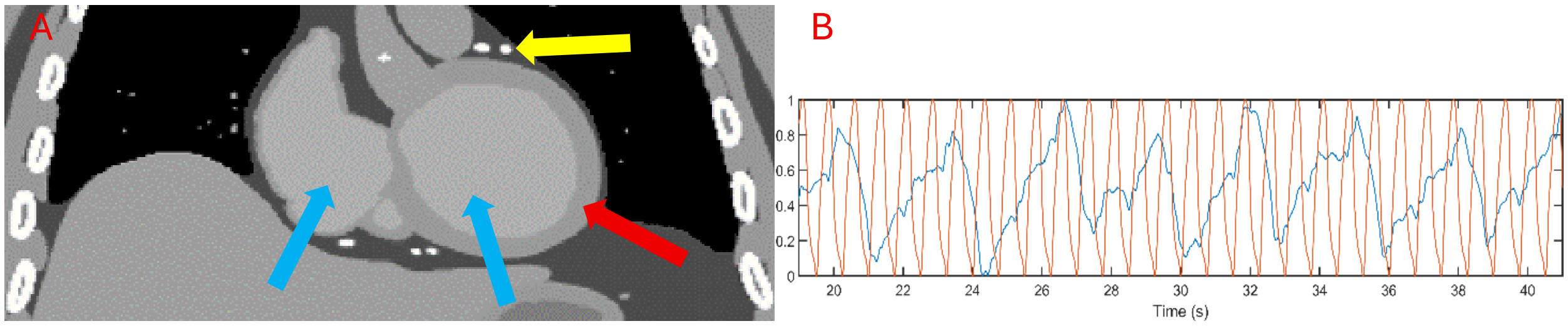

*Figure 1. A: A coronal frame of the 4D XCAT phantom. The contrast between the blood pool (blue arrow), coronary arteries (yellow arrow), and the myocardium (red arrow) is enhanced. B: Blue curve: realistic respiratory motion curve. Red curve: cardiac motion curve. Y-axis represents the normalized amplitude.*

## 2.2. 5DCT scan and reconstruction

5DCT scans were simulated using the 4D XCAT digital phantom under clinical CT scanner hardware settings and scanning parameters as shown in Table 1, which was performed by the TIGRE toolbox[21]. The time, angle, slice position, and cardiac and respiratory phases of each projection were recorded. Each projection was added with simulated CT noise, including Poisson-distributed quantum noise (relevant to the photon-counting process) and random electronic (readout) noise. The acquired projections from the 5DCT scan were used to reconstruct the 3DCT volume with target cardiac and respiratory phases.

The acquired projections were binned with 50 bins (5 cardiac × 10 respiratory). We acquired the cardiac bins using the traditional phase-based binning method (since we used periodic cardiac waves), which is commonly used in the clinic. The respiratory bins were acquired by a hybrid phase-and-amplitude-based binning method[9]. The midpoints of the cardiac and respiratory bins determined the target cardiac and respiratory phases.

Because acquired helical projections are sparsely distributed across the joint cardiac–respiratory phase space, we reconstructed each target slice via a 2D scattered barycentric interpolation with the following steps:

1. **Projection selection**: For a specific row of a specific sinogram, projections with the same angle and with slice positions near the specific sinogram position (distance within half of the detector width projected to the isocenter, ~28 mm) were chosen.
2. **Delaunay triangulation**: Each chosen projection has a cardiac phase value and a normalized respiratory amplitude value. Therefore, a 2D sinogram space with cardiac phase and normalized respiratory amplitude as two dimensions was formed, and each data point in the 2D sinogram space represents one chosen projection. 2D Delaunay triangulation was performed on the data points in the 2D sinogram space by MATLAB. The smallest triangle (Figure 2A) including the target was chosen.
3. **Single-slice rebinning**: For each projection corresponding to the three vertices of the chosen triangle, a rebinned row was acquired by rebinning the helical cone beam projection data to pseudo-fan-beam data[17].
4. **Barycentric interpolation**: We chose it for computational simplicity and stability with sparse, scattered phase-space samples. All rebinned rows were used to perform a 2D barycentric interpolation (with cardiac phase values and normalized respiratory amplitudes as the two source variables) to acquire the row corresponding to the target. The explicit mathematical equation for this interpolation is: For any query point $p = (x_q, y_q)$ lying strictly inside a triangle, let the three vertices of that triangle be $p_1, p_2, p_3$, and let their corresponding data values (rebinned rows) be $v_1, v_2, v_3$. The interpolated row $f(p)$ is computed as:

$$f(p) = \lambda_1 v_1 + \lambda_2 v_2 + \lambda_3 v_3 \quad (1)$$

where $\lambda_1, \lambda_2, \lambda_3$ are the barycentric coordinates of $p$ with respect to the triangle. They satisfy: $p = \lambda_1 p_1 + \lambda_2 p_2 + \lambda_3 p_3$ and $\lambda_1 + \lambda_2 + \lambda_3 = 1$. This barycentric scheme is equivalent to a first-order (piecewise-linear) approximation of the sinogram row as a function of cardiac phase and respiratory amplitude within each triangle: the interpolated row is assumed to vary linearly between the three vertex projections. It may introduce interpolation error where cardiac or respiratory motion is rapid, non-linear, or irregular between the selected vertices, which is a limitation partially addressed by the local weighted linear regression approach described later for cases where the underlying phase relationship is not well-approximated by a single triangle (e.g., near cardiac phase boundaries).

5. **Filtered back projection**: After computing projection data at all projection angles for a 2D slice, a filtered back projection[22] was performed to reconstruct the 2D slice.

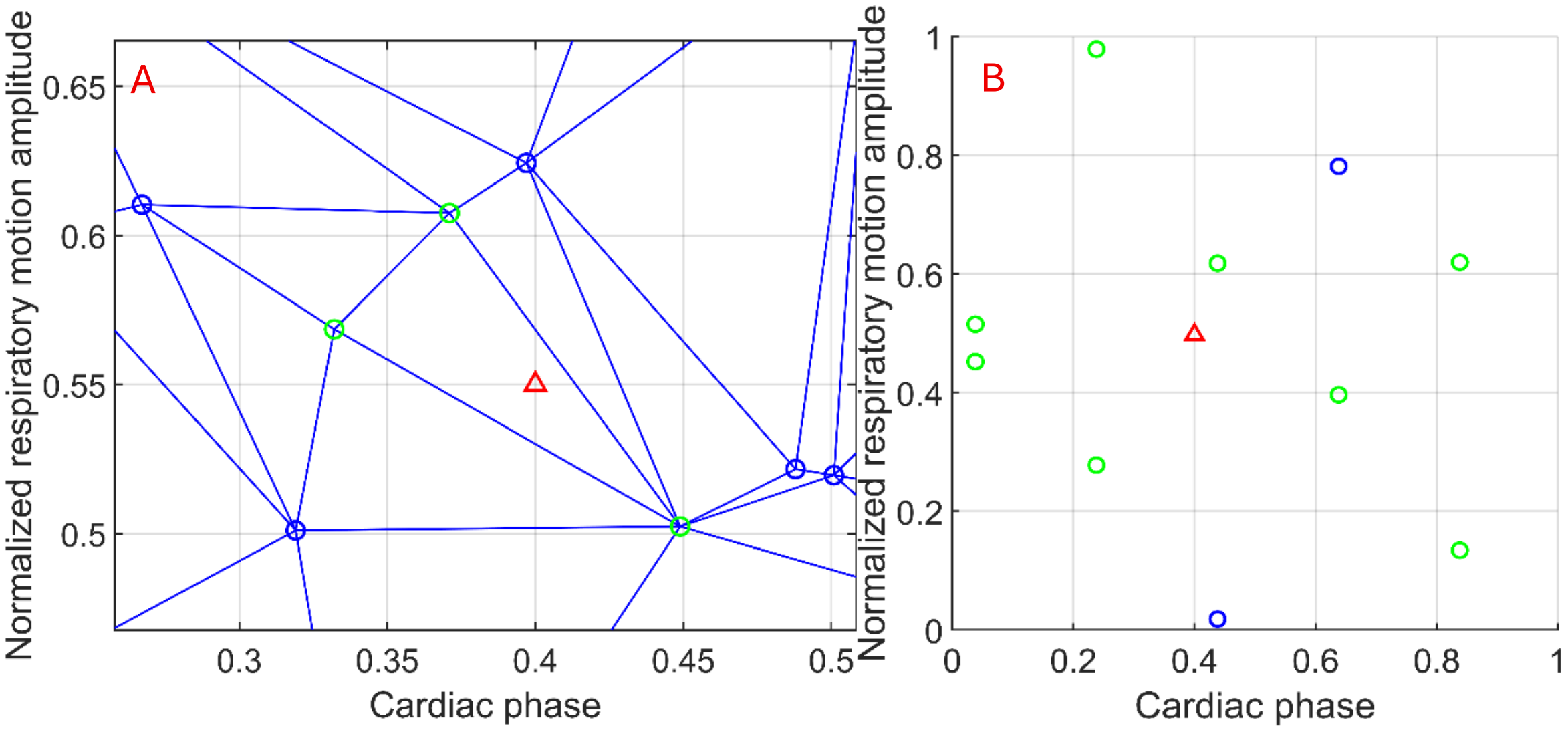

*Figure 2. A: 2D scattered barycentric interpolation for a specific angle at a specific sinogram position. The red triangle represents the target cardiac phases and respiratory amplitudes for the barycentric interpolation. Green and blue circles represent the projections selected for the specific angle and sinogram position. The three green circles represent the projections chosen for single-slice rebinning[17] and linear barycentric interpolation to acquire the sinogram row corresponding to the red triangle. B: 2D scattered local linear interpolation for a specific angle at a specific sinogram position. Green and blue circles*

*represent the projections selected for the specific angle and sinogram position. The green circles are the five nearest points to the target (red triangle).*

If the target cardiac phase is near the boundary (e.g., 00% or 80%), no triangle may include the target after triangulation for some angles and sinogram positions. To solve this problem, the data points in the 2D sinogram-space were translated to the left and right sides (along the cardiac phase dimension) of the original data points to form triangles that include the target near the boundary. This repetition is reasonable since the cardiac motion is periodic. However, we did not perform similar repetition along the respiratory dimension (Y-axis) since it represents the respiratory amplitude[9]. In this case, we assigned target respiratory amplitudes not close to the boundary (e.g., respiratory amplitude 00% is not amplitude 0, it is a bit larger than 0).

If there was still no triangle that included the target after triangulation due to a lack of projection data, we would find the 2 nearest data points to the target, rebin the corresponding projections, project the target point to the line connecting the 2 data points and interpolate the rebinned rows to the projection point. The resulting interpolated sinogram row was treated as the sinogram row of the target. If one of the 2 nearest data points was too far away from the target (e.g., >0.3 in the cardiac dimension), we would treat the rebinned row of the nearest data point as the row of the target.

The cardiac bins were acquired by the phase-based binning method commonly used in the clinic. The cardiac phase values were proportional to time. However, the cardiac motion amplitude is more proportional to the cosine or sine of the time, given that the cardiac motion is circular. If we replace the cardiac phase values with the cosine of the values in the 2D scattered barycentric interpolation, the situation where no triangle may include the target after triangulation will be more common. If we still use the repetition method mentioned before, there will be more fake triangles (two same vertices created by translation) used for barycentric interpolation. To handle the problem, another method called 2D scattered local linear interpolation was developed with the following steps:

1. **Projection selection**: same as step 1 of the 2D scattered barycentric interpolation.
2. **Compute similarity weights:** For cardiac phase (circular), a von Mises-like kernel is used to compute the cardiac weight:

$$w_i^{card} = exp(\kappa(cos(2\pi(c_i - c^*)) - 1)) \quad (2)$$

Where $c_i$ and $c^*$ are the cardiac phase values of the selected data points and the target, $\kappa =$

1 in our experience.

The respiratory weight was computed by a Gaussian kernel:

$$w_i^{resp} = exp(-\frac{(a_i - a^*)^2}{2\sigma_a^2}) \quad (3)$$

Where $a_i$ and $a^*$ are the respiratory motion amplitudes of the selected data points and the target, $\sigma_a$=0.15. The combined similarity weights were computed by $w = w_i^{card} \cdot w_i^{resp}$.

3. **Single-slice rebinning**: On the 2D cardiac phases and respiratory amplitude plane, eight data points around the target were chosen. This was done by creating four quadrants with the target points as the origin, then selecting the two nearest data points to the target for each quadrant (Figure 2B). Next, perform the same rebinning process as the barycentric interpolation for each corresponding projection of the eight data points. If the selected points were fewer than 3 due to a lack of data, the rebinned row of the nearest data point to the target would be treated as the row of the target.
4. **Local weighted linear regression and approximation:** Using the eight selected points $(c_i, a_i)$ and their associated rebinned rows $y_i$, we fit a first-order Taylor expansion of the sinogram rebinned rows $y$ around the target $(c^*, a^*)$:

$$y \approx \beta_0 + \beta_1(c - c^*) + \beta_2(a - a^*) \quad (4)$$

Then we minimize the following weighted least-squares normal equation:

$$min \sum_{j=1}^{8} w_j(y_j - (\beta_0 + \beta_1 \Delta c_j + \beta_2 \Delta a_j)^2 \quad (5)$$

Where $\Delta c_j$ is the signed circular difference of $c_i - c^*$, and $\Delta a_j = a_i - a^*$. The solution is:

$$\beta = \begin{bmatrix} \beta_0 \\ \beta_1 \\ \beta_2 \end{bmatrix} = (X^T W X)^{-1} X^T W y \quad (6)$$

Where $X = \begin{bmatrix} 1 & \Delta c_1 & \Delta a_1 \\ \vdots & \vdots & \vdots \\ 1 & \Delta c_8 & \Delta a_8 \end{bmatrix}$, $W = diag(w_1, \dots, w_8)$, and $y = [y_1, \dots, y_8]^T$. Since we are interested in the target point where the approximate value $y' = \beta_0 = e_1(X^T W X)^{-1} X^T W y$, $e_1 = [1, 0, 0]$. The local linear weights $l^T = e_1(X^T W X)^{-1} X^T W$. The sinogram row at the target point could be approximated by $y' = l^T y$.
5. **Filtered back projection**: same as step 5 of the 2D scattered barycentric interpolation.

In addition to 5DCT reconstruction, we also performed respiratory 4DCT (r4DCT) reconstruction with projection data at pitch = 0.1, a common clinical pitch value. Since each projection corresponds to a respiratory amplitude value, we used 1D interpolation to interpolate rebinned rows to the target respiratory amplitude.

For quantitative evaluation, we generated 50 phases (5 cardiac phases × 10 respiratory amplitudes) of ground-truth CT volumes and 5DCT volumes with pitch values of 0.02 and 0.1 using 2D scattered barycentric interpolation and 2D scattered local linear interpolation, respectively. The ground truth CT volumes were generated by scanning and reconstructing the XCAT phantom volumes at target phases. We evaluated 5DCT reconstruction quality using the mean absolute error (MAE) between the 5DCT and ground-truth CT.

# 3. Results

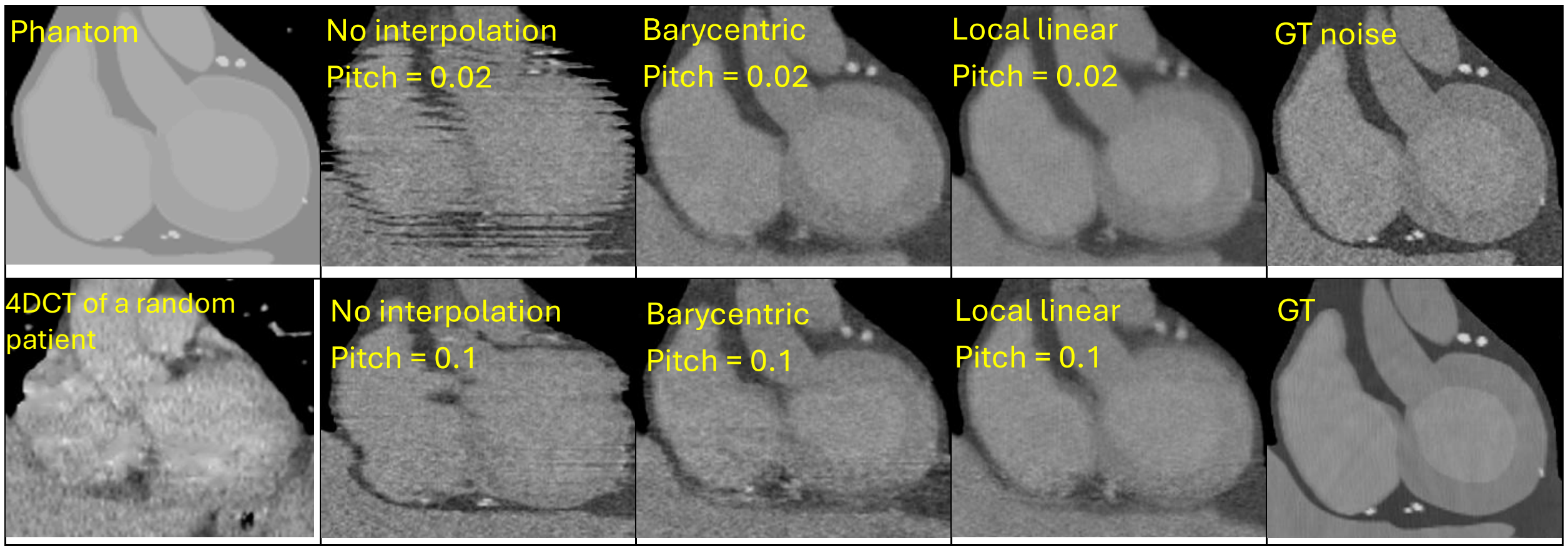

*Figure 3. The XCAT phantom, random patient 4DCT, ground truth (GT) CT with and without simulated noise, and CT volumes reconstructed without and with two interpolation methods with pitch values = 0.02 and 0.1 for a cardiac and a respiratory phase.*

Figure 3 shows the 5DCT reconstruction result in the coronal view without and with 2D scattered barycentric interpolation and local linear interpolation, and the corresponding XCAT phantom and ground truth CT for a cardiac and a respiratory phase. All the reconstructed CTs have simulated noise. Figure 3 shows that the reconstructed CTs have severe misalignment artifacts in the axial direction if reconstructed without our interpolation methods. These artifacts were caused by cardiac and respiratory motion and mostly disappeared if reconstructed with our interpolation methods. For the CT volumes reconstructed with interpolation methods, a lower pitch leads to fewer artifacts. Local linear interpolation produces smoother CT images compared to barycentric interpolation.

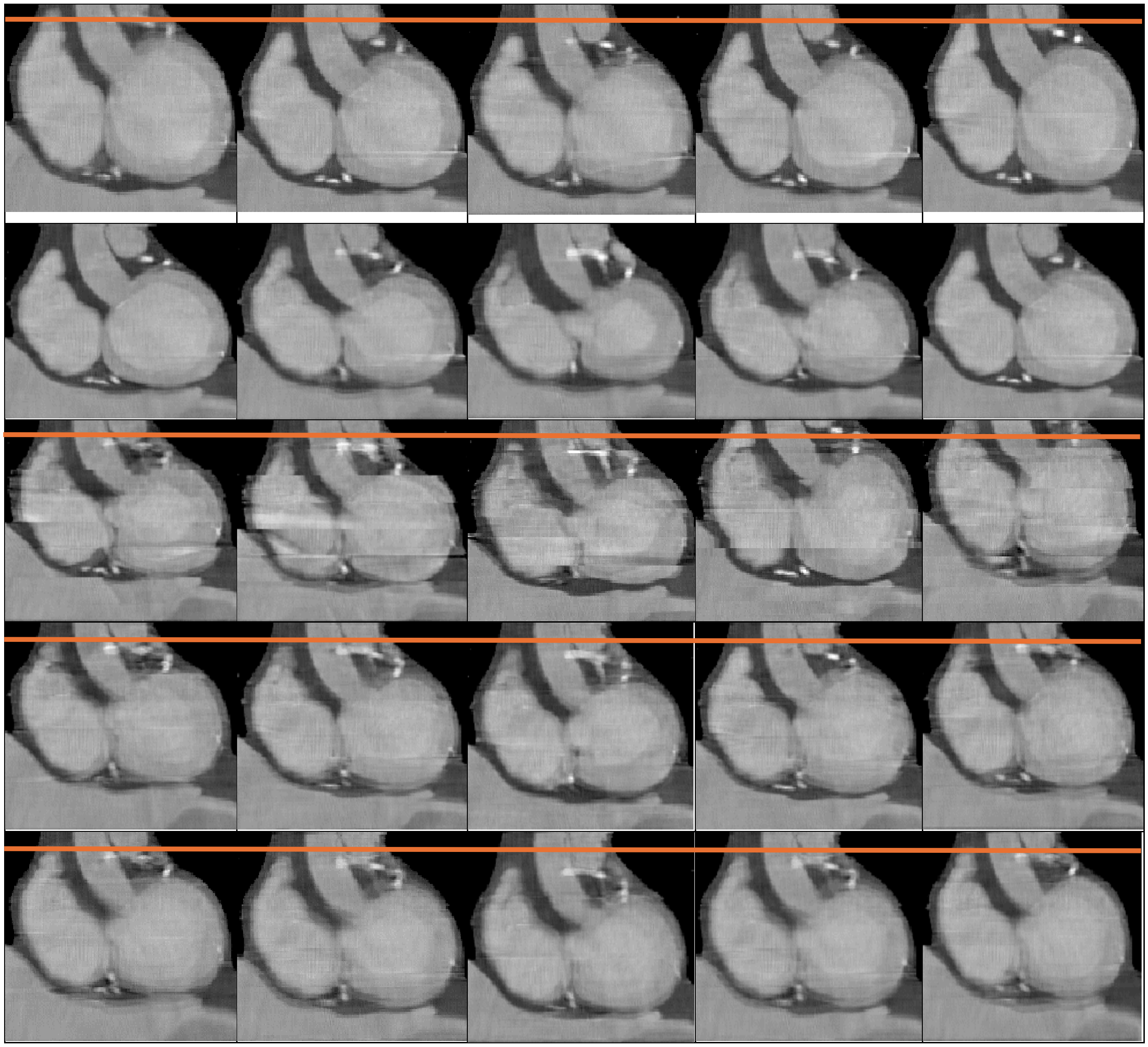

*Figure 4. From top to bottom: First row: five phases of the respiratory-motion-only part of the 5DCT reconstructed with Barycentric interpolation with pitch = 0.02. Second row: five phases of the cardiac-motion-only part of the 5DCT reconstructed with Barycentric interpolation with pitch = 0.02. Third row: five phases of the reconstructed respiratory 4DCT without interpolation with pitch = 0.1. Fourth row: five phases of the reconstructed respiratory 4DCT with interpolation with pitch = 0.1. Fifth row: five phases of the respiratory-motion-only part of the 5DCT reconstructed with Barycentric interpolation with pitch = 0.1.*

Figure 4 shows the reconstructed 5DCT and 4DCT with motion information without simulated noise. For the 5DCT reconstructed with Barycentric interpolation with pitch = 0.02 (first and second rows), the respiratory-motion-only part does not contain obvious cardiac motion artifacts, and the cardiac-motion-only part does not contain obvious respiratory motion artifacts. The contrast between the blood pool and the myocardium and between the coronary arteries and their nearby tissues is clear. For the r4DCT results (third and fourth rows), the reconstructed images without interpolation show obvious cardiac and respiratory motion artifacts. If using interpolation in the respiratory amplitude dimension, the artifacts were greatly reduced but still contained cardiac motion artifacts (see the fourth row; the blood pool volume and myocardium shape were changing across different phases). For the respiratory-motion-only part of the 5DCT reconstructed with Barycentric interpolation with pitch = 0.1 (fifth row), the image contained more streaking artifacts compared to lower-pitch results (first row) due to a lack of data. However, the image contained fewer cardiac motion artifacts (the blood pool volume and myocardium shape were less varied) compared to the r4DCT results (fourth row). This indicates that 5DCT reconstruction could be performed with clinical r4DCT settings to acquire a respiratory-motion-only 4DCT with fewer cardiac motion artifacts. The supporting document shows additional reconstruction results.

| | Barycentric, large FOV | Barycentric, small FOV | Local linear, large FOV | Local linear, small FOV |
|---|---|---|---|---|
| Pitch = 0.02 | 28.2±53.2 | 24.0±46.0 | 21.8±46.8 | 22.9±45.9 |
| Pitch = 0.10 | 37.9±78.2 | 34.7±68.3 | 33.5±84.2 | 38.1±89.3 |

*Table 2. Mean absolute errors under different fields of view (FOV) between the ground truth CT and the reconstructed 5DCT from different reconstruction methods and pitch values. All CT volumes in the computation did not have simulated noise. The unit is HU.*

Table 2 shows the average MAEs and standard deviations (STDs) between the reconstructed 5DCTs and the corresponding ground truth CTs. The MAEs and STDs were computed by calculating the absolute voxel-wise difference between the reconstructed 5DCT volume and the corresponding ground-truth CT volume for each of the 50 target phases (5 cardiac × 10 respiratory), then taking the mean and STD of these differences across all voxels and all phases. Figure 5 shows the size of the FOVs.

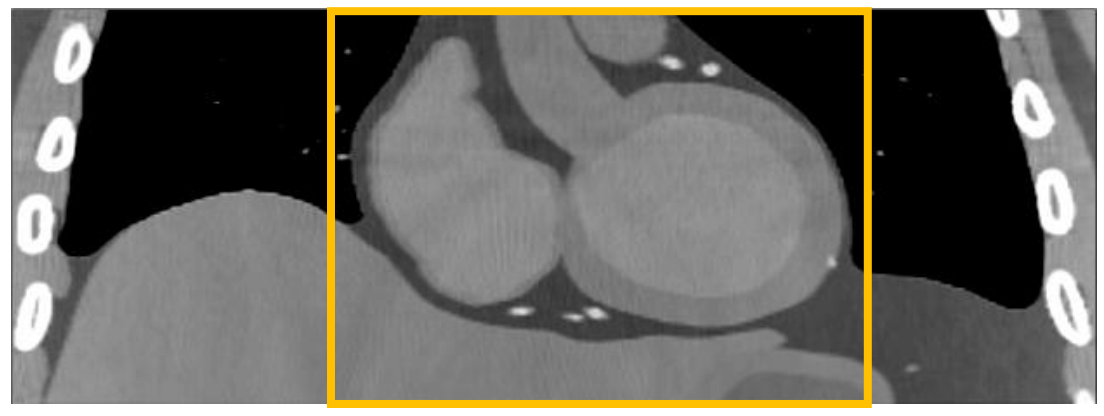

*Figure 5. The whole figure shows the large FOV, and the yellow box shows the small FOV in Table 2. The HU values of the heart are between 1000 and 1100, and the HU values of the lung are around 180.*

For all reconstruction methods and FOVs, the MAE and STD are substantially lower at a pitch of 0.02 compared to 0.10. This is expected, as the lower pitch yields more projections per phase bin, enabling more accurate interpolation and reducing artifacts. The local linear interpolation method consistently yields slightly lower or comparable MAEs than barycentric interpolation at both pitches, with differences more pronounced at pitch 0.02 (e.g., large FOV: 21.8 vs. 28.2 HU). This improvement is likely due to local linear interpolation's ability to handle circular cardiac phases via the von Mises kernel and its adaptive weighting scheme, which better accommodates the periodic nature of cardiac motion and mitigates boundary issues where triangles are not well-defined.

Overall, Table 2 demonstrates that lower pitch and local linear interpolation both contribute to reduced reconstruction errors, with the combination of pitch 0.02 and local linear interpolation providing the best quantitative performance. The relatively large standard deviations indicate that reconstruction quality varies across the 50 phases, underscoring the challenge of reconstructing regions with rapid or complex motion.

# 4. Discussion

This study demonstrates that 5DCT reconstruction using sinogram-space interpolation substantially reduces the axial misalignment artifacts that otherwise result from applying conventional 3D/4D reconstruction to helically acquired cardiorespiratory-resolved projection data (Figure 3). By resolving both cardiac and respiratory phases simultaneously, the resulting cardiac-motion-only and respiratory-motion-only volumes each isolate one physiological motion source while suppressing the other (Figure 4), which is not achievable with conventional cardiac-gated or respiratory-gated 4DCT alone.

Comparing the two interpolation strategies, the results in Table 2 show that local linear interpolation consistently achieved lower or comparable MAE relative to barycentric interpolation, with the advantage most pronounced at the lower pitch. This is consistent with the two methods' underlying assumptions: barycentric interpolation treats cardiac phase as a linear (non-circular) variable and relies on exactly three enclosing data points, which becomes unstable near the 0%/80% cardiac phase boundary and requires an artificial data-repetition workaround to guarantee triangle formation. Local linear interpolation instead models the circular nature of cardiac phase directly through the von Mises–like kernel and draws on eight quadrant-balanced neighboring points rather than three, giving the regression access to phase-space support on both sides of the target in each dimension. This likely explains both its lower quantitative error and the smoother visual appearance reported for its reconstructions (Figure 3), though the larger effective neighborhood also means local linear interpolation trades some spatial/temporal specificity for stability, which is a bias-variance tradeoff that would be worth quantifying directly (e.g., edge-sharpness or motion-blur metrics) in future work, since MAE alone does not distinguish a smoothed-but-biased reconstruction from a sharp-but-noisy one. Barycentric interpolation remains attractive where computational simplicity is prioritized and where the target phase is not near a circular boundary, since it involves a single triangulation and closed-form solve rather than a weighted least-squares fit at every sinogram row.

The relatively large STDs in Table 2 relative to the mean MAEs indicate that reconstruction accuracy is not uniform across the 50 phase combinations; error is likely concentrated in phases coinciding with rapid cardiac motion (the MAE and STD are 22.5±45.0 and 41.3±111.7 at cardiac phases with rapid motion for local linear, small FOV, and pitch = 0.02 and 0.10. The average STD for pitch = 0.10 is much larger than 89.3 which is for all cardiac phases), where the local linearity assumption underlying both interpolation methods is most likely to be violated. This has direct relevance to STAR treatment planning, since it suggests that positional uncertainty, and therefore the safety margin needed around a moving cardiac target, may itself be phase-dependent rather than uniform across the cardiac cycle, a distinction not captured by conventional single-margin approaches.

Notably, the 5DCT pipeline itself can be run at standard clinical pitch (0.1), and the respiratory-only part of that 5DCT still showed fewer cardiac-motion artifacts than a conventional r4DCT reconstructed from the same data (Figure 4, fourth vs. fifth row). This suggests a practical near-term use: our 5DCT reconstruction could be applied to the projection data already acquired for routine r4DCT, and its respiratory-only output used in place of the conventional r4DCT to improve motion characterization for STAR and other thoracic sites (e.g., lung, breast) without any acquisition change. The lower pitch (0.02) used for our best overall 5DCT results is a separate tradeoff, since it requires longer scan times and higher dose than routine protocols.

This study has the following limitations. The 4D XCAT phantom is built from averaged, population-based anatomical models with mathematically defined organ motion. It doesn't capture patient-specific anatomical variability and structural heart disease. A phantom validated on idealized anatomy may not reflect the motion behavior of the diseased hearts this pipeline is ultimately meant to image. While the study added breathing irregularities to the respiratory curve to test robustness, the underlying cardiac motion remains periodic and driven by a single parametric waveform. Real patients can have irregular or variable heartbeats (the very condition motivating STAR), which the phantom's periodic cardiac model doesn't represent. In addition, because the "true" volumes and phase labels come directly from the phantom's own generative model, accuracy assessed against this ground truth may not transfer to real acquisitions, where true anatomy and phase are never perfectly known.

## 5. Conclusion

We developed and evaluated a 5DCT imaging pipeline that resolves cardiac and respiratory motion jointly using sinogram-space interpolation. We proved that 5DCT is theoretically feasible. Local linear interpolation with a circular cardiac-phase kernel outperformed barycentric interpolation, especially at low pitch, and both methods removed the motion artifacts present in uncorrected reconstructions. The pipeline also improved respiratory 4DCT quality even at standard clinical pitch, offering a near-term way to reduce cardiac-motion artifacts without changing acquisition protocols. Future work should validate this approach on real patient data and evaluate its impact on treatment margins for STAR.

## 6. Acknowledgement